\documentclass[twocolumn,5p,times]{article}
\usepackage{graphicx}
\usepackage{lscape,graphicx}
\usepackage{hhline}
\usepackage{multicol}
\usepackage{colordvi}
\usepackage{dcolumn}
\usepackage[fleqn]{amsmath}
\usepackage[mathscr]{euscript}
\usepackage[usenames, dvipsnames]{color}
\usepackage[usenames,dvipsnames,svgnames,table]{xcolor}
\usepackage[linktocpage]{hyperref}
\usepackage[a4paper,bindingoffset=0.2in,left=0.5in,right=0.5in,top=1.0in,bottom=1.0in,footskip=0.25in]{geometry}
\usepackage{caption}
\usepackage{subcaption}
\usepackage{slashed}
\newcommand*{\doi}[1]{\href{http://dx.doi.org/#1}{doi: #1}}
\usepackage[numbers]{natbib}
\makeatletter
\renewcommand{\section}{\@startsection{section}{1}{\z@}{-3.5ex \@plus -1ex \@minus -.2ex}{1.3ex \@plus.2ex}{\normalfont\small\bfseries\boldmath}}
\makeatother

\makeatletter
\renewcommand{\subsection}{\@startsection{subsection}{2}{\z@}{-3.5ex \@plus -1ex \@minus -.2ex}{1.3ex \@plus.2ex}{\normalfont\small\bfseries\boldmath}}
\makeatother

\makeatletter
\renewcommand{\subsubsection}{\@startsection{subsubsection}{3}{\z@}{-3.5ex \@plus -1ex \@minus -.2ex}{1.3ex \@plus.2ex}{\normalfont\small\bfseries\boldmath}}
\makeatother

\makeatletter
\everymath{\if@display\else\thickmuskip=0mu plus 0mu\fi}
\makeatother

\title{\large \bf  {\tt ee$\in$MC}: Comments on Asymmetries in QED }
\date{}

\author{\normalsize Ian M. Nugent$^{*}$ \\ \normalsize Victoria, B.C., Canada}

\begin{document}
\twocolumn[
  \begin{@twocolumnfalse}
    \maketitle
\begin{abstract}
In the Quantum Electrodynamics process $e^{+}e^{-}\to l^{+}l^{-}(n\gamma)$, there are two well known angular asymmetries in the $cos(\theta)$ and the $cos(\theta^{*})$ distributions. 
In this paper, the QED angular asymmetry related to 
the $cos(\theta^{*})$ distribution is investigated in terms of the Dirac propagator and the associated boundary conditions from which the Dirac propagator is constructed  
and the potential implications are examined.  
\end{abstract}
\end{@twocolumnfalse}
]
\renewcommand{\thefootnote}{\fnsymbol{footnote}}
\footnotetext[1]{Corresponding Author \\ \indent   \ \ {\it Email:} inugent.physics@outlook.com}
\renewcommand{\thefootnote}{\arabic{footnote}}
\section{Introduction}

In Quantum Electrodynamics (QED) processes, the higher order emission of hard photons directly influences the angular dependence in the differential cross-section. Consequently, for the 
$e^{+}e^{-}\to\mu^{+}\mu^{-}(\gamma)$  and $e^{+}e^{-}\to\tau^{+}\tau^{-}(\gamma)$ processes, this is related to the angular asymmetry between the outgoing charged leptons 
$cos(\theta)$ \cite{2006257,Renton:1990td,Peskin:1995ev,Halzen:1984mc,Ryder:1985wq,Mandl:1985bg,Griffiths:1987tj} and $cos(\theta^{*})$ \cite{BaBar:2015onb}, asymmetry in the
angle between the outgoing lepton and the radiated hard photon in the center-of-mass frame of the outgoing lepton pair. 
Higher order contributions, in particular, Feynman Diagrams with $k=2,...$ internal photon exchanges further modify the $cos(\theta)$ asymmetry \cite{Campanario:2013uea}\footnote{These 
are the most significant terms in the infinite perturbative series.}, where the virtual and soft-photon 
contributions are most significant. This is in contrast to the $cos(\theta^{*})$ angular distributions, where the asymmetry originates from the inclusion of Feynman Diagrams that contain $n=1$ or more 
real hard photon emissions. In {\it ee$\in$MC} \cite{Nugent:2022ayu}, each of the hard matrix elements, ${{\mathcal M}}_{n}^{k}$ for $n$ real hard photon emissions and $k$ internal photon exchanges, 
are explicitly calculated without approximations from the Gamma-Matrices and Dirac spinors using an object-oriented structure. This allows for the investigation of the angular asymmetry $cos(\theta^{*})$ 
in terms of the treatment of the Dirac propagator and the associated boundary conditions. 

\section{Overview of {\tt ee$\in$MC} Formalism}

The {\tt ee$\in$MC} \cite{Nugent:2022ayu} Monte-Carlo generator, is a stand-alone software program which contains the random-number generation \cite{MT32,MT64Tab,MT64F2,xorshift,Knuth:1981}, the phase-space
generation \cite{Nugent:2022ayu,Nugent:2022eeinmc} based on the \cite{Byckling:1969} algorithm  modified to include embedded importance sampling \cite{Lopes:2006,Gelman:2014} 
and the theoretical models for the QED processes $e^{-}e^{+}\to\mu^{+}\mu^{-}(n\gamma)$, $e^{-}e^{+}\to\tau^{+}\tau^{-}(n\gamma)$,
 $e^{-}e^{+}\to hadrons(n\gamma)$ and  $\tau$ lepton decays.
The cross-section for the QED processes $e^{-}e^{+}\to\mu^{+}\mu^{-}(n\gamma)$, $e^{-}e^{+}\to\tau^{+}\tau^{-}(n\gamma)$ and $e^{-}e^{+}\to hadrons(n\gamma)$ is constructed within the
Yennie-Frautschi-Suura (YFS) Exponentiation Formalism \cite{Yennie:1961} for the infra-red subtraction

\begin{equation}
\resizebox{0.375\textwidth}{!}{$
d\sigma= \frac{\sum_{n=0}^{\infty} \Pi_{i=0}^{n}\Pi_{j=0}^{j< i}F\left(Y_{i,j}^{\mathcal{O}(\alpha)}(P_{i}^{\mu},P_{j}^{\mu})\right)|\sum_{k=1}^{\infty}\bar{{\mathcal M}}_{n}^{k}|^{2} dPS_{n}}{4(|\vec{P}_{e^{-}\
}|E_{e^{+}}+E_{e^{-}}|\vec{P}_{e^{+}}|)}
$}
\label{eq:YSR}
\end{equation}

\noindent where $F(x)$ is some functional form representing the resummation of all permutations for soft or virtual photon exchanges. For the Initial and Final YFS multiplicative subtraction, the 
function form $F(x)$ is the standard exponential YSF Form-Factor for the  Yennie-Frautschi-Suura  calculation \cite{Yennie:1961};
the {\tt KK2F} approximation  \cite{kk2f}; and the Sudakov Form-Factor \cite{Peskin:1995ev}. For the Full LO calculation from \cite{Schwinger:1998} applying corrections from
\cite{Peskin:1995ev,Nugent:2022ayu,Smith1994117}, the $F(x)$ is the product of the exponential Form-Factor determined from  \cite{Schwinger:1998} with the Coulomb potential factored out into a separate
multiplicative resummation series, the Sommerfeld-Sakharov resummation factor \cite{Smith1994117}.  The hard matrix elements, $\bar{{\mathcal M}}_{n}^{k}$, is determined with the spin-average-sum for an 
arbitrary Initial-State spin configuration \cite{Nugent:2022tt} and is explicitly determined from the Feynman calculus corresponding to each Feynman Diagrams using an object-orientated representation of the
Gamma-Matrices and Dirac Spinors \cite{Nugent:2022ayu}.  Ward's Identity \cite{Mandl:1985bg} is applied to incorporate the renormalization through the running of electromagnetic 
coupling constant \cite{Nugent:2022ayu,Jegerlehner,Sturm_2013}. Details on the simulation of the $\tau$ decays can be found in \cite{Nugent:2022ayu,Nugent:2023eie}.

\section{Dirac Propagator Formalism \label{sec:DiracPop}}

It is well known that the Dirac propagator can be described in terms of the retarded Green's function for a free particle with retarded boundary conditions \cite{Peskin:1995ev},

\begin{equation}
\resizebox{0.3\textwidth}{!}{$
 G=\imath<0|\{\psi_{a}(x^{\prime})\bar{\psi}_{b}(x)\}|0>\theta(x_{(0)}^{\prime}-x_{(0)}).
$}
\label{eq:DiracPropDef}
\end{equation}

\begin{figure*}[tbp]
\begin{center}
\resizebox{500pt}{178pt}{
    \includegraphics{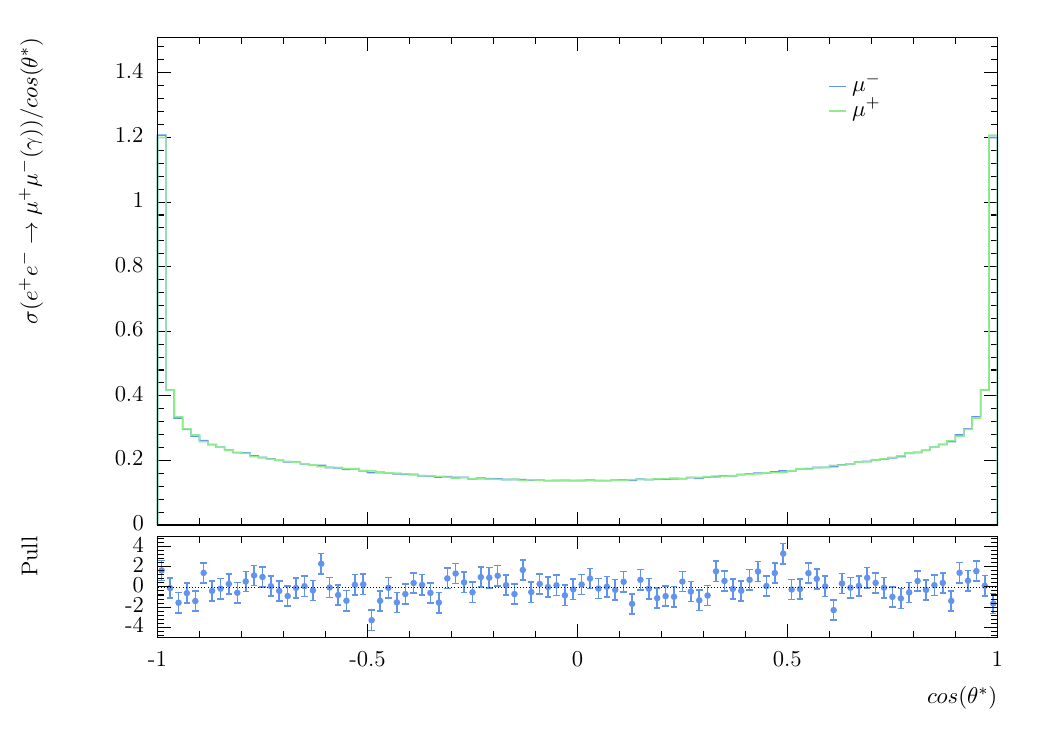}
    \includegraphics{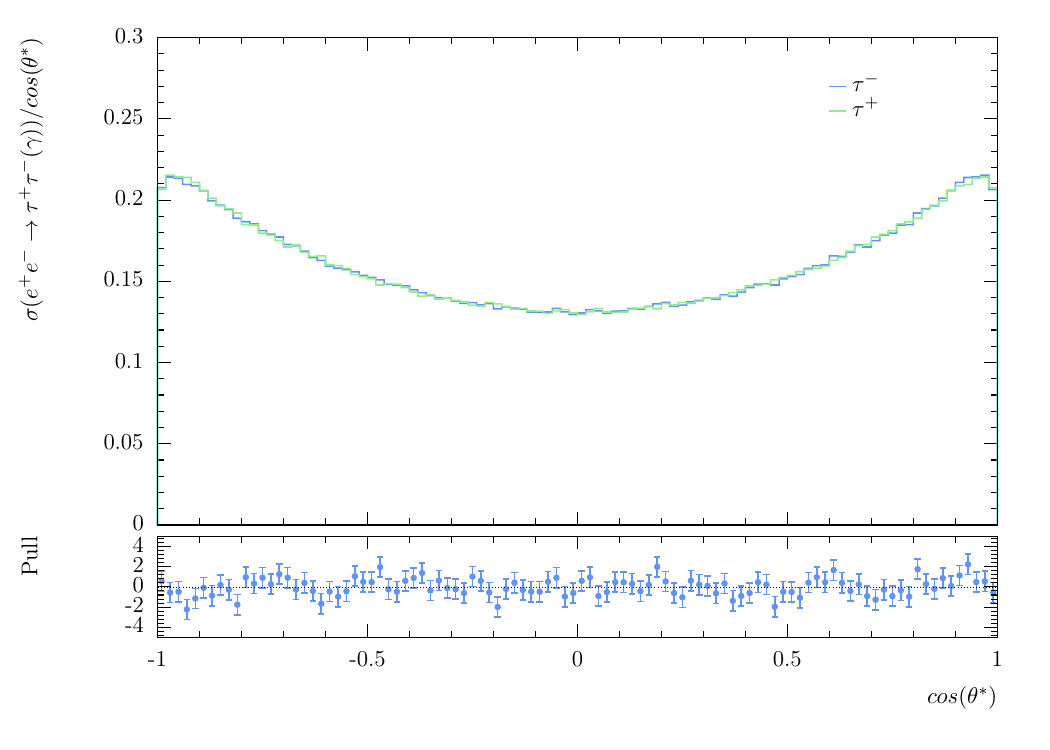}
  }
\resizebox{500pt}{178pt}{
    \includegraphics{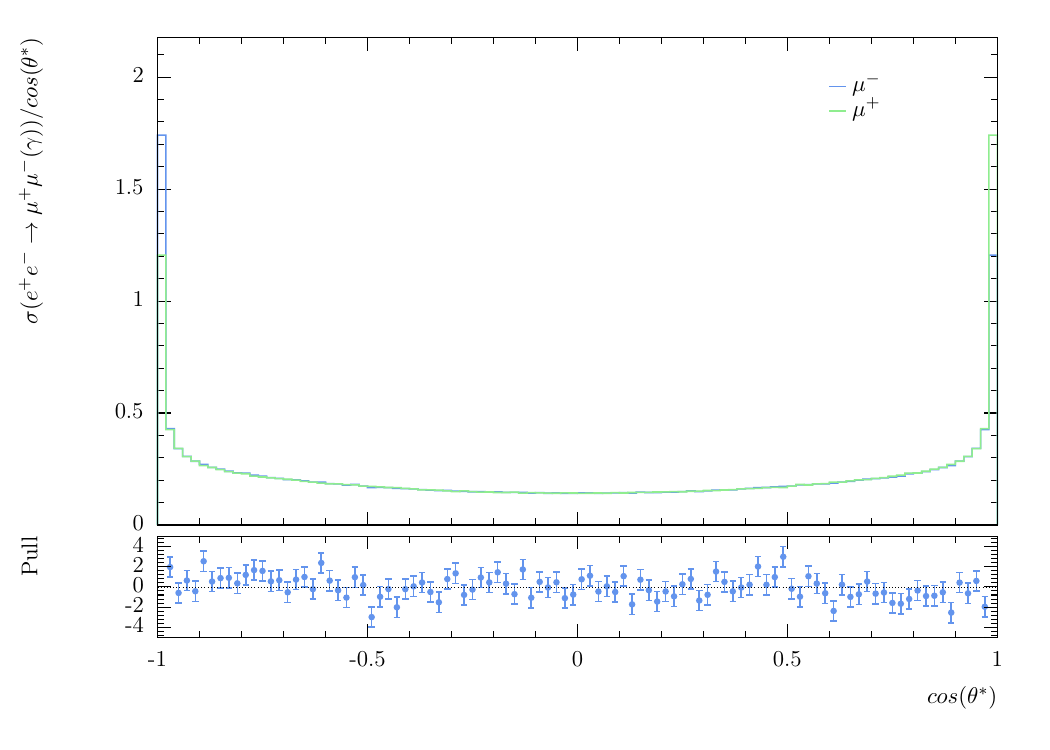}
    \includegraphics{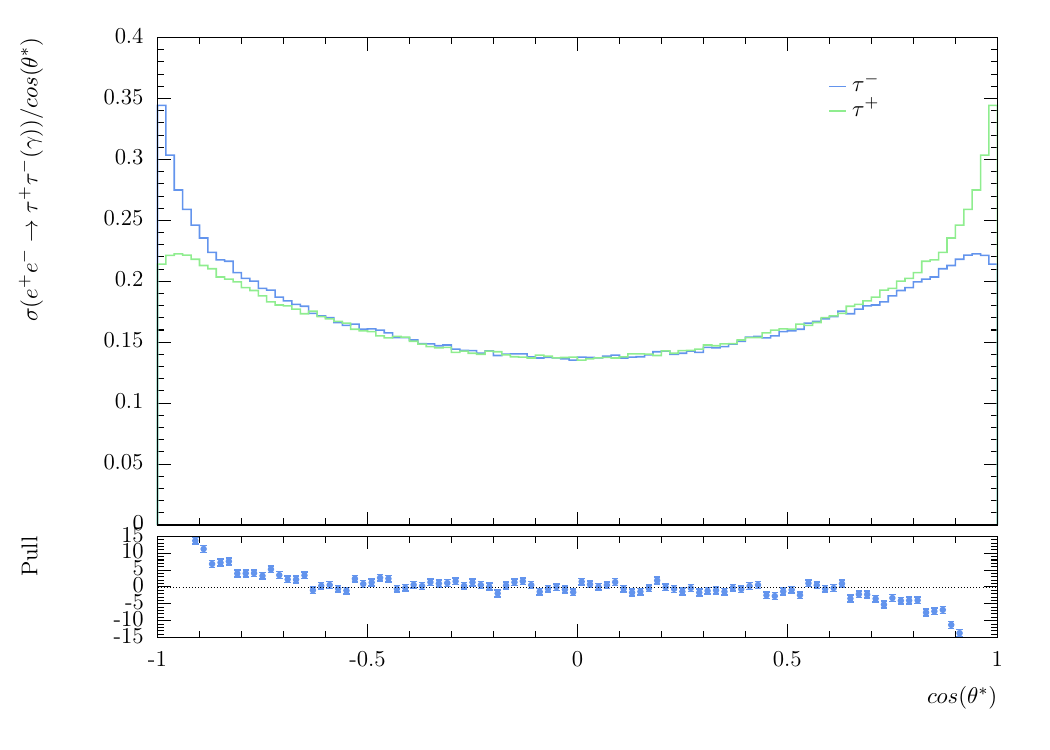}
 }
\end{center}
  \caption{The $cos(\theta^{*})$ distribution for Born+LO determined with 
the Dirac propagator constructed from the Feynman boundary conditions (top) and
the retarded Dirac propagator (bottom) for the
$e^{+}e^{-}\to\mu^{+}\mu^{-}(\gamma)$  (left)  and  $e^{+}e^{-}\to\tau^{+}\tau^{-}(\gamma)$ (right)
QED interactions for a center-of-mass energy of $10.58GeV/c^{2}$ and a soft-photon cut-off of $E_{l}=1MeV$. The blue line represents the $cos(\theta^{*})$ angle between the photon and $l^{-}$
in the rest frame of the outgoing lepton pair, while the green line represents the $cos(\theta^{*})$ angle between the photon and $l^{+}$
in the rest frame of the outgoing lepton pair.  
\label{fig:costhetastar} }
\end{figure*}

\noindent From which it follows that: 
\begin{equation}
\resizebox{0.25\textwidth}{!}{$  
 \left(\imath\gamma^{\mu}\frac{\partial}{\partial x_{\mu}^{\prime}}-m\right)G(x^{\prime}-x)=\delta^{4}(x^{\prime}-x).
$}
\label{eq:DiracPropDef}
\end{equation}
\noindent Using the Fourier Transform between the momentum and the coordinate space\cite{Renton:1990td,Peskin:1995ev,Halzen:1984mc}, 
this can trivially be represented as:
\begin{equation}
\resizebox{0.18\textwidth}{!}{$                                                                                                                                                                                   
G(p)=\frac{1}{\slashed{p}-m}=\frac{\slashed{p}+m}{p^{2}-m^{2}}
$}
\label{eq:DiracProp}
\end{equation}
\noindent \cite{Renton:1990td,Peskin:1995ev,Halzen:1984mc}. It is the retarded Green's function, Equation \ref{eq:DiracProp}, which is generally used to
 describe the propagator for each internal fermion line in 
perturbation theory \cite{Renton:1990td,Peskin:1995ev,Halzen:1984mc,Ryder:1985wq,Mandl:1985bg,Griffiths:1987tj}. 
In contrast, the Dirac propagator defined with the Feynman Boundary conditions is written as:
\begin{equation}
\resizebox{0.425\textwidth}{!}{$
  \begin{array}{lll}
G(x^{\prime}-x)&=&\int \frac{d^{4}p}{(2\pi)^{4}}e^{-\imath p\cdot (x^{\prime}-x)}G(p)\\
&=&-\imath \int\frac{d^{3}p e^{\imath p\cdot (x^{\prime}-x)}}{(2\pi)^{3}2E}\left(\left(\gamma^{0}E-\gamma\cdot{\mathbf p}+m\right)e^{-\imath E(t^{\prime}-t)}\theta(t^{\prime}-t) \right. \\
&& \left. + \left(-\gamma^{0}E-\gamma\cdot{\mathbf p}+m\right)e^{-\imath E (t-t^{\prime})}\theta(t-t^{\prime})\right) \\
&=&-\imath \int\frac{d^{3}p}{(2\pi)^{3}2E}\left(\sum_{s}u_{a}^{s}(p)\bar{u}_{b}^{s}e^{-\imath p\cdot (x^{\prime}-x)}\theta(t^{\prime}-t) \right. \\
&& \left.- \sum_{s}v_{a}^{s}(p)\bar{v}_{b}^{s}e^{-\imath p\cdot (x-x^{\prime})}\theta(t-t^{\prime})\right)\\
&=&-\imath \int\frac{d^{3}p}{(2\pi)^{3}}\left(\frac{m}{E}\Lambda_{+}e^{-\imath p\cdot (x^{\prime}-x)}\theta(t^{\prime}-t) \right. \\
&& \left.+ \frac{m}{E}\Lambda_{-}e^{-\imath p\cdot (x-x^{\prime})}\theta(t-t^{\prime})\right)\\
&=&-\imath<0|T\{\psi_{a}(x^{\prime})\bar{\psi}_{b}(x)\}|0>.
\end{array}
$}
\label{eq:DiracPropFull}
\end{equation}
\noindent \cite{Renton:1990td,Peskin:1995ev} $\Lambda_{+}$ and $\Lambda_{-}$ are the standard positive and negative projection operators for the Dirac spinors,
\begin{equation}
\begin{array}{lr}
\Lambda_{+}=\frac{\left(\slashed{p}+m\right)}{2m}, & \Lambda_{-}=\frac{\left(-\slashed{p}+m\right)}{2m} \\
\end{array}
\label{eq:SpinorProjOp2}
\end{equation}
\noindent \cite{Renton:1990td}. This is a time-ordered solution, where the integration contour for $t^{\prime}-t>0$ is in the lower half of the plane and corresponds to the 
positive energy solution while the integration contour for $t^{\prime}-t<0$  is in the upper half plane and corresponds to the negative energy solution. The $\slashed{p}-m$ and $\slashed{p}+m$ 
factors project out the positive and negative contribution to the propagator. Then, given that $(\sum_{s}u_{a}^{s}(p)\bar{u}_{b}^{s}=\left(\slashed{p}+m\right)$ 
and $\sum_{s}v_{a}^{s}(p)\bar{v}_{b}^{s}=\left(\slashed{p}-m\right)$, and Equation \ref{eq:DiracPropFull}
it follows that $G_{+}=\frac{\slashed{p}+m}{p^{2}-m^{2}}$ and $G_{-}=\frac{\slashed{p}-m}{p^{2}-m^{2}}$ for the positive and negative energy states going forward and backward in time respectively.
From the derivation of the Feynman calculus, it can be seen that before applying Wick's Theorem to obtain the non-vanishing propagators and vertices, the expectation values in the 
$s$-matrix for the perturbative expansion must be time-ordered \cite{Renton:1990td}. Na\"ively, this implies that the Dirac propagators should also be time-ordered. Therefore, 
if one applies the Green's function with Feynman boundary conditions directly to the positive energy fermion states ($u^{s}(p)$) and negative energy fermion states 
($v^{s}(p)$), taking into account the time-ordered direction of the particle/anti-particle states one obtains interesting results in terms of the angular asymmetries in QED\footnote{At 
this stage we would like to remind the reader of the Feynman rules for writing down the fermion line in the QED process $f^{+}f^{-}\to f^{\prime +}f^{\prime -}$. More specifically, going right to left 
for incoming particles, one starts with the positive energy state going forward in time to the electro-magnetic vertex and then proceeds backwards in time for the anti-fermion line. Similarly, for
the outgoing particles one starts with the anti-fermion line going backwards in time to the QED vertex, and then proceeds forward in time for the outgoing fermion line.}\footnote{This formulation 
of the Dirac propagator also has implications for the box-diagram terms. We noted that in other processes, for example the $K_{L}$ $K_{S}$ mass difference \cite{Renton:1990td}, 
that Feynman boundary conditions for the Dirac propagator are consistent with known results.  }.
Figure  \ref{fig:costhetastar} presents radiative Born plus LO cross-section dependence on the angle between the emitted $\gamma$ and the outgoing lepton ($\mu^{-}$ or $\tau^{-}$) in the center-of-mass frame 
for the outgoing lepton-pair, $cos(\theta^{*})$. The reported asymmetry in the
$cos(\theta^{*})$ distribution \cite{BaBar:2015onb,MC:2010}, is reproduced by the Dirac propagator corresponding to the retarded Green's function solution. 
The asymmetry is most
significant at $cos(\theta^{*})=\pm1$. This region is removed in many other MC generators. The $cos(\theta^{*})$ angular asymmetry is more strongly peaked in the  $e^{+}e^{-}\to\mu^{+}\mu^{-}(\gamma)$ process
while the $e^{+}e^{-}\to\tau^{+}\tau^{-}(\gamma)$ process is more spread out, a consequence of the larger $\tau$ mass. 
When the time-ordered propagator is applied for the
corresponding particle and anti-particle respectively, no asymmetry is observed in the $cos(\theta^{*})$ distribution. This suggests that the $cos(\theta^{*})$ asymmetry in QED is
directly related to the choice of the Dirac propagator and the application in the perturbation theory. 

\section{Conclusion}

The $cos(\theta^{*})$ asymmetries in the $e^{+}e^{-}\to\mu^{+}\mu^{-}(\gamma)$  and $e^{+}e^{-}\to\tau^{+}\tau^{-}(\gamma)$ QED interactions were investigated in terms of the choice of boundary conditions
applied in the formulation of the Dirac propagators. The differential cross-section is symmetric in $cos(\theta^{*})$ for the Feynman boundary conditions which incorporate the time-ordering 
of the positive and negative energy Dirac states when applied to the Born and LO simulation, while the retarded Green's function for the Dirac propagator has a clear asymmetry. The asymmetry is most 
significant at $cos(\theta^{*})=\pm1$, where due to the mass of the $\tau$ lepton is more visible away from the angular boundaries for the $e^{+}e^{-}\to\tau^{+}\tau^{-}(\gamma)$. 
 We argue that applying the Feynman boundary conditions for the Dirac propagator is more consistent with 
the time-ordering from which the perturbative $s$-matrix is constructed. 

\section*{Acknowledgement}
GCC Version 4.8.5 was used for compilation and the plots are generated using the external program GNUPlot \cite{gnuplot4.2}.

\footnotesize
\bibliography{paper}
\normalsize

\end{document}